%% file: ms.tex
\PassOptionsToPackage{table}{xcolor}
\documentclass[acmtog,authorversion=true]{acmart}

\usepackage{booktabs} 

\usepackage[normalem]{ulem}
\usepackage{color}
\usepackage{soul}
\usepackage{graphicx}
\usepackage{caption}
\usepackage[export]{adjustbox}
\usepackage{subcaption}
\usepackage{epstopdf}
\usepackage{color}
\usepackage{xspace}
\usepackage[final]{pdfpages}
\usepackage{textcomp}
\usepackage{gensymb}
\usepackage{pdfpages}
\usepackage{grffile}
\usepackage[inline]{enumitem}
\usepackage{comment} 
\hypersetup{draft}

\setcitestyle{square}
\input{src_defines}
\begin{document}

	\input{00_title}

	\author{Julio Marco}	\authornote{Part of this work was done during Julio Marco's internship at Microsoft Research Asia.}\author{Quercus Hernandez}\author{Adolfo Mu\~{n}oz}
	\affiliation{%
		\institution{Universidad de Zaragoza, I3A}
	}
	\email{}

	\author{Yue Dong}
	\affiliation{%
		\institution{Microsoft Research Asia}
	}

	\author{Adrian Jarabo}
	\affiliation{%
		\institution{Universidad de Zaragoza, I3A}
	}

	\author{Min H. Kim}
	\affiliation{%
		\institution{KAIST}
	}

	\author{Xin Tong}
	\affiliation{%
		\institution{Microsoft Research Asia}
	}

	\author{Diego Gutierrez}
	\affiliation{%
		\institution{Universidad de Zaragoza, I3A}
	}
						
	\renewcommand{\shortauthors}{Marco et al.}

	%
	%

	\setcopyright{acmcopyright}
	\acmJournal{TOG}
	\acmYear{2017}\acmVolume{36}\acmNumber{6}\acmArticle{219}\acmMonth{11} 
	\acmDOI{10.1145/3130800.3130884}
	
	
	\begin{teaserfigure}
	\centering
	\includegraphics[width=0.95\textwidth]{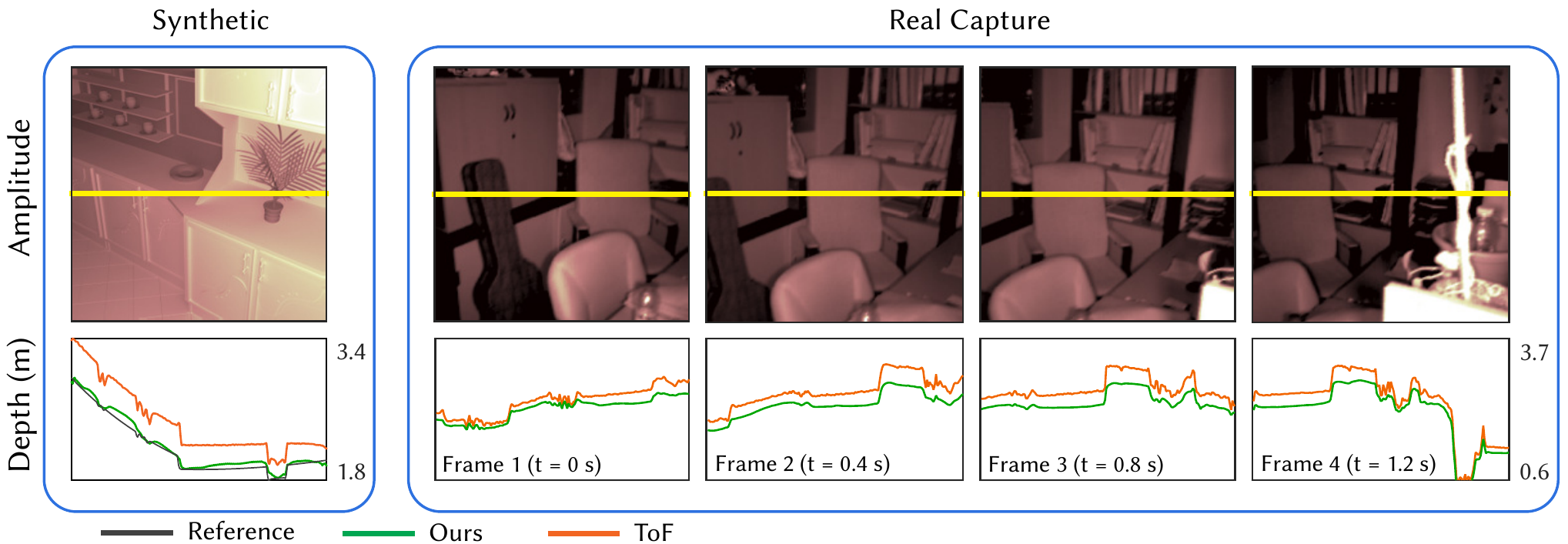}
	\caption{We propose a convolutional approach to correct multipath interference (MPI) errors, in real-time and on complex scenes captured with off-the-shelf, time-of-flight (\ToF) cameras. We introduce a novel two-stage learning scheme: we first pre-initialize our network to learn real depth representations, and then re-train these to learn MPI corrections from synthetic data. The leftmost image shows one of our synthetic validation scenes: our method (green) provides a better fit to  ground truth data than \ToF depth from the camera. In the  sequence to the right, we illustrate how our system allows for real-time MPI compensation (10 milliseconds per frame) in real scenes captured with \ToF devices, preserving temporal consistency. Please refer to the supplemental video for more results. }
	\label{fig:teaser}
	\end{teaserfigure}
	
	\input{01_abstract}

	\maketitle

	\input{10_introduction}

	\input{20_related_work}
	\input{30_problem_statement}

	\input{31_our_approach}
	\input{40_training_set}\input{50_learning_design}

	\input{60_results}

\input{70_conclusion}

	\input{99_acks}

	\appendix
	\input{80_light_transport}
	\input{90_depth_stats}

	\bibliographystyle{ACM-Reference-Format}
	\bibliography{bibliography}
	
\end{document}

%% file: src_defines.tex
\definecolor{red}{rgb}{0.8,0,0}
\definecolor{purered}{rgb}{1,0,0}
\definecolor{darkred}{rgb}{0.6,0,0}
\definecolor{green}{rgb}{0.0,0.5,0}
\definecolor{blue}{rgb}{0,0,0.75}
\definecolor{darkblue}{rgb}{0,0,0.55}
\definecolor{orange}{rgb}{0.9,0.3,0.1}
\definecolor{purple}{rgb}{0.6,0.0,0.6}
\definecolor{cyan}{rgb}{0.0,0.7,0.7}
\definecolor{darkgray}{rgb}{0.4,0.4,0.4}
\definecolor{bronze}{rgb}{0.8, 0.5, 0.2}
\definecolor{dorange}{rgb}{0.8, 0.2, 0.0}

\newcommand{\diegoc}[1]{\textcolor{red}{\emph{(\textbf{Diego:} #1)}}}
\newcommand{\diego}[1]{\diegoc{#1}}

\newcommand{\MIN}[1]{\textcolor{red}{\emph{(\textbf{Min:} #1)}}}

\newcommand{\review}[1]{#1}

\newcommand{\Fig}[1]{Figure~\ref{fig:#1}}
\newcommand{\Figs}[2]{Figures~\ref{fig:#1} and \ref{fig:#2}}
\newcommand{\Tab}[1]{Table~\ref{tab:#1}}
\newcommand{\Eq}[1]{Equation~\eqref{eq:#1}}

\newcommand{\Sec}[1]{Section~\ref{sec:#1}}

\newcommand{\Apx}[1]{Appendix~\ref{ap:#1}}

\newcommand{\pixel}{p}
\renewcommand{\exp}[1]{e^{#1}}
\newcommand{\sImaginary}{j}

\newcommand{\vect}[1]{\mathbf{#1}}           
\newcommand{\vpath}[1]{\overline{#1}} 
\newcommand{\pixelpath}{\vpath{\pixel}}
\newcommand{\pixelpathrange}{P}

\newcommand{\ToFSensorFreq}{f_{\omega_R}}

\newcommand{\ToFAmplitude}{x} 
\newcommand{\ToFPhase}{\phi} 
\newcommand{\ToFDepth}{z} 

\newcommand{\sCorr}{\vect{c}} 
\newcommand{\sCorrApp}{\hat{\vect{c}}} 

\newcommand{\RGBD}{RGB-D\xspace}

\newcommand{\sTransportMatrix}{\vect{T}} 

\newcommand{\ToF}{ToF\xspace}
\newcommand{\x}{$\times$}	

%% file: 00_title.tex
\title{DeepToF: Off-the-Shelf Real-Time Correction of Multipath Interference in Time-of-Flight Imaging}


%% file: 01_abstract.tex
\begin{abstract}
Time-of-flight (\ToF) imaging has become a widespread technique for depth estimation, allowing affordable off-the-shelf cameras to provide depth maps in real time. 
However, multipath interference (MPI) resulting from indirect illumination significantly degrades the captured depth. 
Most previous works have tried to solve this problem by means of complex hardware modifications or costly computations. 
In this work, we avoid these approaches and propose a new technique to correct errors in depth caused by MPI, which requires no camera modifications and takes just 10 milliseconds per frame.
%
Our observations about the nature of MPI suggest that most of its information is available in image space; this allows us to formulate the depth imaging process as a spatially-varying convolution and use a convolutional neural network to correct MPI errors. 
Since the input and output data present similar structure, we base our network on an autoencoder, which we train in two stages. First, we use the encoder (convolution filters) to learn a suitable basis to represent MPI-corrupted depth images; then, we train the decoder (deconvolution filters) to correct depth 
from synthetic scenes, generated by using a physically-based, time-resolved renderer. %
This approach allows us to tackle a key problem in \ToF, the lack of ground-truth data, by using a large-scale captured training set with MPI-corrupted depth to train the encoder, and a smaller synthetic training set with ground truth depth to train the decoder stage of the network. 
We demonstrate and validate our method on both synthetic and real complex scenarios, using an off-the-shelf \ToF camera, and with only the captured, incorrect  depth as input.

\end{abstract}

%% file: 10_introduction.tex
\section{Introduction}
Time-of-flight (\ToF) imaging, and in particular continuous-wave \ToF cameras, have become a standard technique for capturing depth maps. 
Such devices compute depth of visible geometry by emitting modulated infrared light towards the scene, and correlating different phase-shifted measurements at the sensor. However, \ToF devices suffer from multipath intereference (MPI): a single pixel records multiple light reflections, but it is assumed that all light reaching it has followed a direct path (see~\cite{Jarabo2017} for details). This introduces an error on the captured depth, which reduces the applicability of \ToF cameras. 

To compensate for MPI, most previous works leverage additional sources of information, such as coded illumination or multiple modulation frequencies that lead to different phase-shifts, from which indirect light might be disambiguated. This requires either hardware changes (e.g. modifying the built-in illumination, or using  sensors that can handle multiple modulated frequencies), or multiple passes with a standard \ToF camera. 
Other single-modulation approaches simulate an estimation of the ground-truth light transport, then compensate the captured MPI using information from such simulation. Such approaches, while working with any out-of-the-box \ToF sensor, typically require several minutes for a single frame, and might lead to errors when the simulation is not accurate. 
%

Our work aims to lift these limitations: We present a novel technique to correct MPI using an \textit{unmodified, off-the-shelf camera}, with a \textit{single frequency}, and in \textit{real time}. A key observation is that, since both the camera and the infrared emitter are co-located and share almost the same visibility frustrum, most of the MPI information is actually present in image space. Furthermore, from the discretized geometry given by a depth map, light transport at each pixel can only be estimated as a linear combination (with unknown weights) of the contributions from the rest of the pixels. This linear process can be represented as a spatially-varying convolution in image space, with unknown convolution filters. This motivates our design of a convolutional neural network (CNN) to obtain such filters. 

However, suitable \ToF datasets that include depth with MPI and its corresponding ground-truth reference do not exist, and capturing such dataset is not possible with current devices. To overcome this, we synthesize such data using an existing physically-based transient light transport renderer (\Sec{datageneration}), extending it with a \ToF camera model. Using this model, we introduce MPI in the simulated depth estimation, and compare it with the reference depth. Our network uses the synthetic data for training, takes depth with MPI as input, and returns a corrected depth map. In particular, since the input and output have the same resolution, we design an encoder-decoder network. While amplitude or phase-shifted images could provide additional information to the network, they are often uncalibrated and highly dependent on the device characteristics. We therefore use depth as the only reliable input to our network, providing a real-time solution that is robust even for single-frequency \ToF devices.

This approach introduces two new challenges. First, generating synthetic data is time-consuming, and thus the simulated dataset is unlikely to be large and diverse enough to avoid overfitting. Second, although we carefully analyze our synthetic data to make sure that it is statistically similar to real-world data, it might still be too perfect, lacking for instance subtle differences due to imperfections in the sensor or the emitter.
We address this by leveraging the fact that the input and output depths must be structurally similar, and devising a two-stage training. The first stage is a convolutional autoencoder (CAE) from real-world captured data, which requires no ground-truth reference. This tackles both challenges, since it allows us to use large datasets with real-world imperfect data for training. 
This first stage thus trains the encoding filters of the network as a feature dictionary learned from structural properties of \ToF depth images (\Sec{cae_learning}). 

The second stage provides supervised learning for the regression with the synthetic dataset as reference, which accounts for the effect of MPI (input and output are now different), and feeds the decoder (while the encoder remains unmodified). We treat the effect of MPI as a residue, and therefore model this second stage as supervised residual learning (\Sec{residualnet}). 

We analyze the performance of our approach in synthetic and captured ground-truth data, and compare against previous works, showing favorable results while being significantly faster. Finally, we demonstrate our technique in-the-wild, correcting MPI from depth maps captured by a \ToF camera in real time. In summary, we make the following contributions: 

\begin{itemize}
\item A two-stage training strategy, with a convolutional autoencoder plus a residual learning approach. It leverages statistical knowledge from real captured data with no ground truth, and then compensates the error from synthetic data (which includes ground truth).
\item A synthetic \ToF dataset of scenes sharing similar statistical properties as real-world scenes. For each scene, we provide both MPI-corrupted and ground-truth depth. We believe this data is much needed, and we hope our dataset can help future works. 
\item A trained network that compensates multipath interference from a single \ToF depth image in real time, which outperforms previous algorithms even using such minimal input.
\end{itemize}
\review{Our training dataset and trained network are publicly available online at \url{http://webdiis.unizar.es/~juliom/pubs/2017SIGA-DeepToF/}}

%

%% file: 20_related_work.tex

\section{Related work}

Convolutional neural networks (CNNs) have been widely used for many image-based reconstruction tasks, such as intrinsic images, normal estimation, or depth recovery. Here we only focus on CNN-based depth reconstruction methods that are closely related to our work. We refer to Jarabo et al.'s recent survey ~\cite{Jarabo2017} for a complete overview on transient and \ToF imaging, and to Goodfellow et. al.'s book
~\shortcite{Goodfellow2016deeplearning} for other deep learning techniques and their applications.

\paragraph*{CNN-based depth reconstruction}
A set of methods derive  depth from multi-view images. 
Zbontar et al.~\shortcite{zbontar2015stereo} trained a CNN for computing the matching cost of stereo image pairs. 
Kalantari et. al.~\shortcite{Kalantari2016LVS} exploited a CNN to estimate the disparity between sparse light-field views, and fed the result to another CNN to interpolate the light-field views for novel view synthesis. Different from these multi-view methods, 
we reconstruct a depth image from a single snapshot captured by a monocular \ToF camera.   

\input{22_mpi_table}

Other methods estimate depth from a single RGB image. 
Eigen et. al.~\shortcite{eigen2014depth,eigen2015depth} proposed an end-to-end CNN in which a coarse depth image is first recovered, then progressively refined. In each step, the coarse depth is upsampled and combined with fine scale image features. Based on this approach, several works formulate the generated depth as a conditional random field (CRF), and then refine it with the help of color image segmentation~\cite{Wang2015depth,liu2015depthcrf,li2015depthcrf}, or multi-resolution depth information generated by intermediate CNN layers~\cite{xu2017depthcrf}. Although these methods improve the accuracy of the result, the CRF optimization is expensive and slow. 
Most recently, Su et. al.~\shortcite{Su2017shape} trained a CNN with a synthetic dataset rendered from a large dataset for reconstructing a low-resolution 3D shape from a single RGB or \RGBD image. 
Different from these methods, we do not rely on additional sources of information. Moreover, we have also developed an efficient scheme that combines both unlabeled real \ToF images and labeled synthetic data for CNN training. This can efficiently generate a full-resolution depth map at a rate of up to 100 frames/second. 

\paragraph*{Multipath interference}
Several works take advantage of inputs with several amplitudes and phase images through multiple modulation frequencies.
This input can be translated into multipath interference correction through optimization~\cite{dorrington2011separating,freedman2014sra}, closed-form solutions with inverse attenuation polynomials~\cite{godbaz2012closedform}, spectral methods~\cite{kirmani2013spumic,feigin2016kinect}, sparse regularization~\cite{bhandari2014resolving}, or through modeling indirect lighting as phasor interactions in frequency space~\cite{gupta2015phasorimaging}.
Although these techniques are efficient for some devices that can capture a few frequencies simultaneously, such as Kinect V2~\cite{feigin2016kinect}, they require multiple passes for single-frequency \ToF cameras.

Other approaches deal with multipath interference by adding or modifying hardware. Wu and colleagues~\shortcite{Wu_IJCV_2014} decompose global light transport into direct, subsurface scattering, and interreflection components, leveraging the extremely high temporal resolution of the femto-photography technique~\cite{Velten_Femto_2013}.  Modified \ToF sensors allow to reconstruct a transient image from multiple frequencies~\cite{heide2013lowbudget,peters2015trigonometric}. Other techniques include custom coding~\cite{kadambi2013coded}, or combining  \ToF sensors with structured light projection~\cite{naik2015light,otoole2014probing}. 
%
In contrast, our method works with just an out-of-the-box phase \ToF sensor, without any modifications.

Some techniques remove multipath interference from a single frequency (a single amplitude and depth image) by estimating light transport from the approximated depth. One of the approximations considers a single indirect diffuse bounce (assuming constant albedo) connecting all pairs of pixels on the scene~\cite{fuchs2010multipath}. This has been later extended to multiple diffuse bounces and multiple albedos, by adding some user input~\cite{fuchs2013compensation}. Last, an optimization algorithm over depth space with path tracing has also been presented~\cite{jimenez2014modeling}. While these approaches manage to compensate multipath interference from input obtained with any \ToF device, they are very time-consuming; moreover, \review{given the sparse input and the assumptions simplifying the underlying light transport model}, they are unable to completely disambiguate indirect light in all cases. In contrast, given even less information (a single depth map, without amplitude) our work is able to compensate multipath interference for varied and complex geometries, with different albedos, and in real time.

\Tab{mpi} summarizes these approaches, and compares them to our method. We list the required input, the variability of the tested scenes (including albedo and geometry) and the execution time. Our method works on a large range of scenes, with a just a single depth map as input, and yields real time performance. 

%% file: 22_mpi_table.tex
\begin{table*}[]
	\caption{Comparison between the different existing techniques that address the problem of multipath interference, including required input, tests shown in the paper (material types and variability of scenes) and execution time. Our approach performs in real time while requiring the most reduced input. Furthermore, we show a greater scene variability in terms of materials and geometries.}
	\label{tab:mpi}
	\newcommand{\good}[1]{\cellcolor{green!25}#1}
	\newcommand{\average}[1]{\cellcolor{yellow!25}#1}
	\newcommand{\bad}[1]{\cellcolor{red!25}#1}
\centering
\footnotesize
\begin{tabular}{lllll}
	\textbf{Work}                   & \textbf{Input}                            & \textbf{Tested materials}&\textbf{Scenes    }& \textbf{Reported time}               \\ \hline
	\cite{fuchs2010multipath}       & \good{Single frequency}                   & \bad{Single albedo}      & \bad{Few planes}  & \bad{$\approx 10$ minutes}          \\                           
	\cite{dorrington2011separating} & \average{Multiple frequencies}             & \good{Multiple albedos}  & \bad{Few planes}  & Unreported                          \\                           
	\cite{godbaz2012closedform}     & \average{Multiple frequencies}             & \good{Multiple albedos}  & \bad{Few planes}  & Unreported                          \\                           
	\cite{fuchs2013compensation}    & \average{Single frequency, user input}    & \good{Multiple albedos}  & \bad{Few planes}  & \bad{$\approx 150$ seconds}         \\  
	\cite{heide2013lowbudget}       & \bad{Multiple freqs. hardware mods.}      & \good{Multiple albedos}            & \average{Several} & \bad{$\approx 90$ seconds}          \\  
	\cite{kadambi2013coded}         & \average{Single freq., custom code}       & \good{Multiple albedos}            & \average{Several} & \average{$\approx 4$ seconds}       \\       
	\cite{kirmani2013spumic}        & \average{Multiple frequencies}            & \bad{Single albedo}      & \bad{Few planes}  & \good{Real time} \\
	\cite{freedman2014sra}          & \average{Multiple frequencies}            & \good{Mult. albedos, specular}&\bad{Few scenes} & \good{$31.2$ ms}                    \\
	\cite{bhandari2014resolving}    & \average{Multiple frequencies}            & \average{Transparency}   & \bad{Few planes}  & Unreported                          \\                           
	\cite{otoole2014probing}        & \bad{Multiple freqs., structured light}   & \good{Multiple albedos}  & \average{Several }& Unreported                          \\                           
	\cite{jimenez2014modeling}      & \good{Single frequency}                   & \good{Multiple albedos}  &\good{Many}        & \bad{Several minutes}               \\                           
	\cite{peters2015trigonometric}  & \average{Multiple frequencies}            & \good{Mult. albedos, specular}&\average{Scene, video}& \good{$18.6$ fps}              \\
	\cite{gupta2015phasorimaging}   & \average{Multiple frequencies}            & \good{Multiple albedos}  & \bad{Few scenes}& Unreported                          \\
	\cite{naik2015light}            & \average{Single freq. structured light}   & \good{Multiple albedos}  & \bad{Few planes}  & \average{"Not real time"}             \\
	\cite{feigin2016kinect}         & \average{Multiple frequencies}            & \average{Transparency}   & \average{Several}  & \average{"Efficient"} \\
	\textbf{Our work}	        & \good{Single depth}                       & \good{Multiple albedos}  & \good{Many}       & \good{10ms}                    \\
	\vspace{0.5mm}
\end{tabular}
\end{table*}

%% file: 30_problem_statement.tex

\section{Problem statement}
\label{sec:problem}
\paragraph*{\ToF depth errors} Using four phase-shifted measurements $\sCorr_{1...4}$, \ToF devices compute the depth $\ToFDepth$ at every pixel $\pixel$ as
\begin{align}
\ToFDepth & = \frac{c\, \ToFPhase }{4\pi \ToFSensorFreq} 
\label{eq:arctandepth} \\
\ToFPhase & = \arctan \left( \frac{\sCorr_{4} - \sCorr_{2}}{\sCorr_{1} - \sCorr_{3}}\right),
\label{eq:arctanangle} 
%
%
\end{align}
where $\ToFSensorFreq$ is the device modulation frequency, $c$ is the speed of light in a vacuum, and $\ToFPhase$ is the phase of the wave reaching a pixel $\pixel$. This model works under the assumption of a single impulse response from the scene, therefore assuming 
\begin{equation}
\sCorr_{i}(\pixel) = \ToFAmplitude(\pixel) \exp{2\pi\,\sImaginary\,\ToFPhase(\pixel) \frac{\ToFSensorFreq}{c} + \theta_i},
\end{equation}
with $\sImaginary=\sqrt{-1}$, $\ToFAmplitude(\pixel)$ the amplitude of the wave, and $\theta_i$ the phase shift of measurement $\sCorr_i$. However, given indirect illumination, the observed pixel may receive light from paths other than single-bounce direct light, which leads to
\begin{equation}
\sCorrApp_{i}(\pixel) = \sCorr_{i}(\pixel) + \int_\pixelpathrange \ToFAmplitude(\pixelpath) \, \exp{2\pi\,\sImaginary\,\ToFPhase(\pixelpath) \frac{\ToFSensorFreq}{c} + \theta_i} d \pixelpath,
\label{eq:pixelmultiple}
\end{equation}
where $\pixelpathrange$ is the space of all the light paths $\pixelpath$ reaching pixel $\pixel$ from more than one bounce; the amplitude $\ToFAmplitude(\pixelpath)$ and phase delay due to light time-of-flight $\ToFPhase(\pixelpath)$ are now functions of the path~$\pixelpath$. 
%
The effect of multiple bounce paths if often ignored in \ToF sensors
obtaining approximate measures 
$\sCorr_{i}(\pixel) = \sCorrApp_{i}(\pixel)$ and therefore leading to a depth estimation error, the multipath interference (MPI).

\begin{figure}[t]
	\centering
	\includegraphics[width=1\columnwidth]{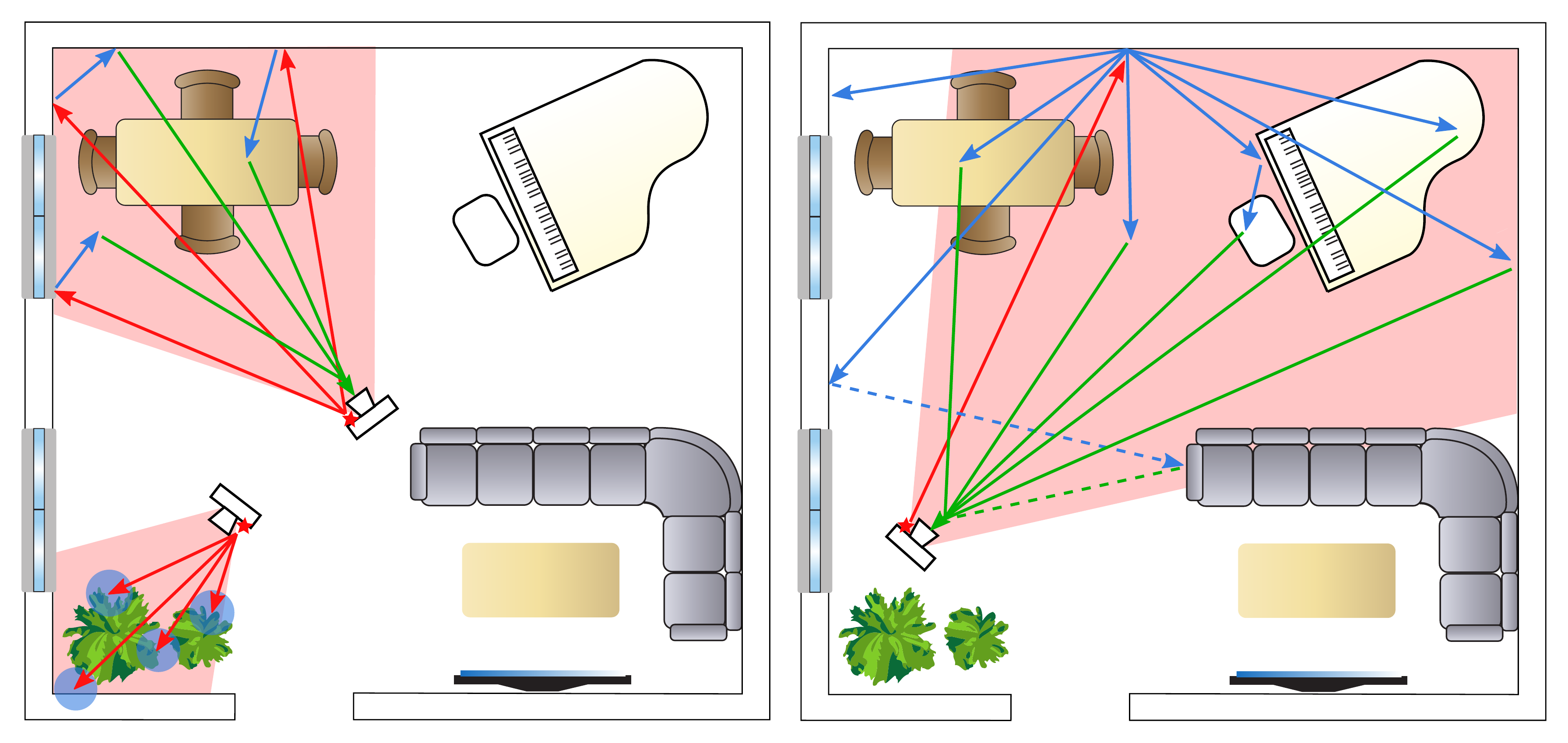}%
	\vspace{-3mm}%
	\caption[]{\ToF MPI at different scales, showing IR emission (red), indirect bounces (blue) and observed radiance (green). Left: Observed second-bounce illumination occurs mostly from reflections on visible geometry due to shared light-camera visibility frustum (camera facing the table), while higher-order bounces usually have a significant impact in the locality of the observed points (e.g. camera facing the plants). Right: Large objects such as a wall may cast a significant indirect component over the whole scene when captured from afar, while longer paths from out-of-sight geometry (discontinuous) create negligible MPI due to attenuation. }
	\label{fig:room_bounces}
	\vspace{-2mm}
\end{figure}
\paragraph*{MPI observations}%
In \ToF range devices, the light source is typically co-located with the camera, sharing a similar visibility frustum. As we illustrate in \Fig{room_bounces} (camera facing the table), this implies that most of the MPI due to second-bounce indirect illumination comes from actual visible geometry. 
Previous works have leveraged this by taking into account only second-bounce illumination~\cite{fuchs2010multipath,dorrington2011separating}, or by ignoring non-visible geometry~\cite{jimenez2014modeling}. Higher-order indirect illumination might come from non-visible geometry, but due to the exponential decay of scattering events and quadratic attenuation with distance, light paths of more than two bounces interfere mainly in the local neighborhood of a point (see \Fig{room_bounces}, camera facing the plants). These observations suggest that most of the information on multipath interference from a scene is available in \textit{image space}, where \Eq{pixelmultiple} is discretized into a summatory and can be modeled as a spatially-varying convolution. Please refer to \Apx{light_transport} for a more detailed derivation of such spatially-varying convolution model.
Last, since the effect of multipath interference does not eliminate major structural features of a depth map, the incorrect depth and the reference depth are structurally similar. 

%% file: 31_our_approach.tex
\section{Our approach}
\label{sec:approach}

Given that MPI can be expressed as a spatially-varying convolution, MPI compensation could be modeled as a set of convolutions and deconvolutions in depth space. This in turn suggests that MPI errors could in principle be solved designing a convolutional neural network (CNN).
Specifically, since incorrect and correct depths are only slightly different (but structurally similar),  using a convolutional autoencoder (CAE) would be a tempting solution. 
A convolutional autoencoder is a powerful tool which takes the same input and output to learn hidden representations of lower-dimensional feature vectors by \textit{unsupervised} learning, resulting in two symmetric networks: an encoder and a decoder.
%
This allows to build a deeper network architecture, and preserves spatial locality when building these representations~\cite{masci2011stacked}. 
The lower-dimensional feature vectors retain the relevant structural information on the input and eliminate existing errors, effectively returning the restored (reference) image. Recently, CAEs have been successfully used in many vision and imaging tasks (e.g.~\cite{Choi2017,Du2016}).

A straight-forward CAE, nevertheless, cannot be applied to our particular problem: as the errors introduced by MPI are highly correlated with the reference depth to be recovered, we need such ground-truth reference for training.
However, a large enough labeled dataset (i.e., pairs of MPI-corrupted depth and its corresponding ground-truth depth), needed for training, does not exist. Although real-world, \ToF depth images are widely available, measuring their ground-truth depth maps is a non-trivial task. On the other hand, rendering time-resolved images from synthetic scenes is extremely time-consuming, and the results would only cover a small portion of real-world scene variations. 

We propose a two-step training scheme to infer our convolutional neural network from both \textit{unlabeled} real depth images, and labeled synthetic depth image pairs (with and without MPI). Our method is inspired by super-resolution methods based on overcomplete dictionaries and sparse coding~\cite{yang:tip:2010}. Figure~\ref{fig:net_arch} shows an overview of our network: We first learn an encoder network as a depth prior through traditional unsupervised CAE training, in which unlabeled real depth images (with unknown errors) are used as both input and output. 
The resulting encoder allows us to obtain lower-dimensional feature vectors of incorrect depth images. In our second step, different from sparse coding where original signals are reconstructed as a linear product of dictionary atoms, we train a decoder that can reconstruct the reference depth map from such feature vectors. To this end, we keep the encoder network unchanged and cascade it with a residual decoder network. The weights of the decoder network are learned from the synthetic depth pairs via supervised CNN training. 

Our network therefore only takes the depth image with MPI as input, and outputs a depth map without the effects of MPI. We do not feed our network with any other \ToF information, such as pixel amplitudes or phase images, because these properties highly depend on specific \ToF camera settings, and are unstable. By using only depth as input, our solution is robust using off-the-shelf \ToF devices that operate with a single frequency.

%
In the following two sections, we first introduce our dataset for training (\Sec{datageneration}), then describe in detail our network and our two-step training scheme (\Sec{learning}). 

%% file: 40_training_set.tex
\section{Training data}
\label{sec:datageneration}

\begin{figure}[t]
	\centering
	\includegraphics[width=\columnwidth, page=1]{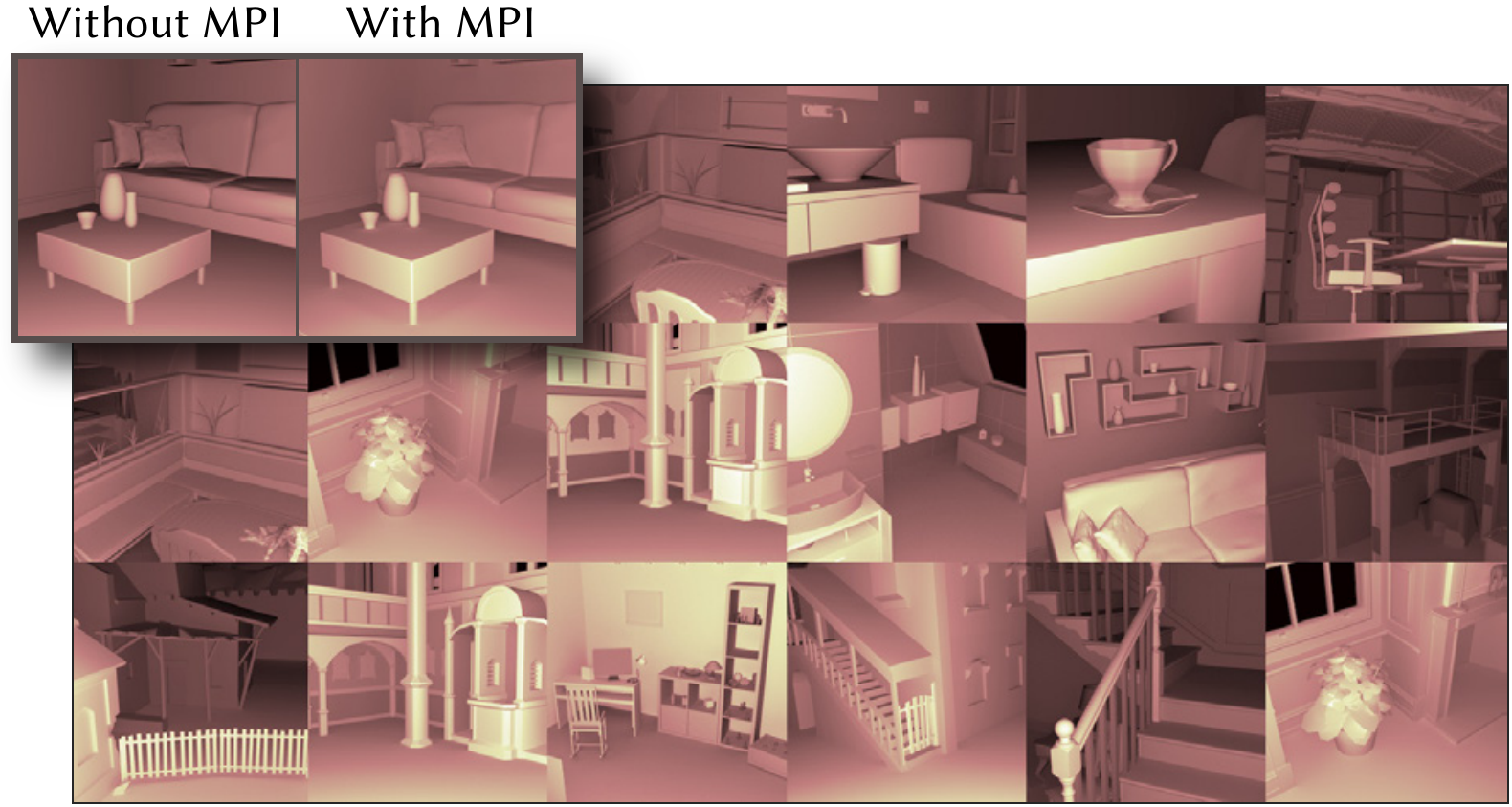}
	\caption{Representative sample of the scenes (amplitude) rendered with the \ToF model to generate depth images with MPI. The top-left image shows the same scene with and without MPI.}
	\label{fig:training_example}
\end{figure}
To obtain accurate pairs of incorrect (due to MPI) and ground-truth reference depth maps, we simulate the response of the \ToF lighting model using the publicly-available, time-resolved bidirectional path tracer of Jarabo et al.~\shortcite{JaraboSIGA14}. This allows us to obtain four phase images [\Eq{arctanangle}], and from those the MPI-distorted depth estimation of a \ToF camera [\Eq{arctandepth}]. 

%
Existing \ToF devices generally use square modulation functions instead of perfect sinusoidal ones; this introduces additional errors on the depth estimation (\emph{wiggling}), although \ToF cameras do account for these errors and compensate them in the final depth image. \review{The wiggling effect and its correction are specific for each camera, and in general information is not provided by manufacturers. }
We therefore use ideal sinusoidal functions to avoid introducing non-MPI-related error sources. Ground-truth depths are straightforward to obtain from simulation.

\paragraph*{Dataset}%
We simulated 25 different scenes with varying materials, using six different albedo combinations between 0.3 and 0.8, and rendered from seven different viewpoints, at 256$\times$256 resolution---similar to what \ToF cameras yield---, computed with up to 20 bounces of indirect lighting. From these we obtained 1050 depth images with MPI (which we flip and rotate to generate a total of 8400) and their respective reference depths. The geometric models of the scenes were obtained from three different free repositories\footnote{https://benedikt-bitterli.me/resources/\\ http://www.blendswap.com/\\ https://free3d.com/}. Some examples  can be seen in \Fig{training_example}. In the remaining of this paper we refer to this as the \textbf{synthetic dataset}.

Training a network from scratch requires a sufficiently large labeled dataset. However, generating it is very time-consuming, so using only synthetic data for our purposes is highly unrealistic. We therefore gather an additional dataset of 6000 \textit{unlabeled}, real depth images (48000 with flips and rotations) from public repositories~\cite{silberman2012indoor,karayev2011category,xiao2013sun3d}, and use them to pre-initialize our network; we refer to this as the \textbf{real dataset}. Learning representations of unlabeled real depths will later improve depth corrections from our smaller synthetic labeled dataset (\Sec{learning}). 

\begin{figure}[t]
	\centering
	\includegraphics[width=\columnwidth]{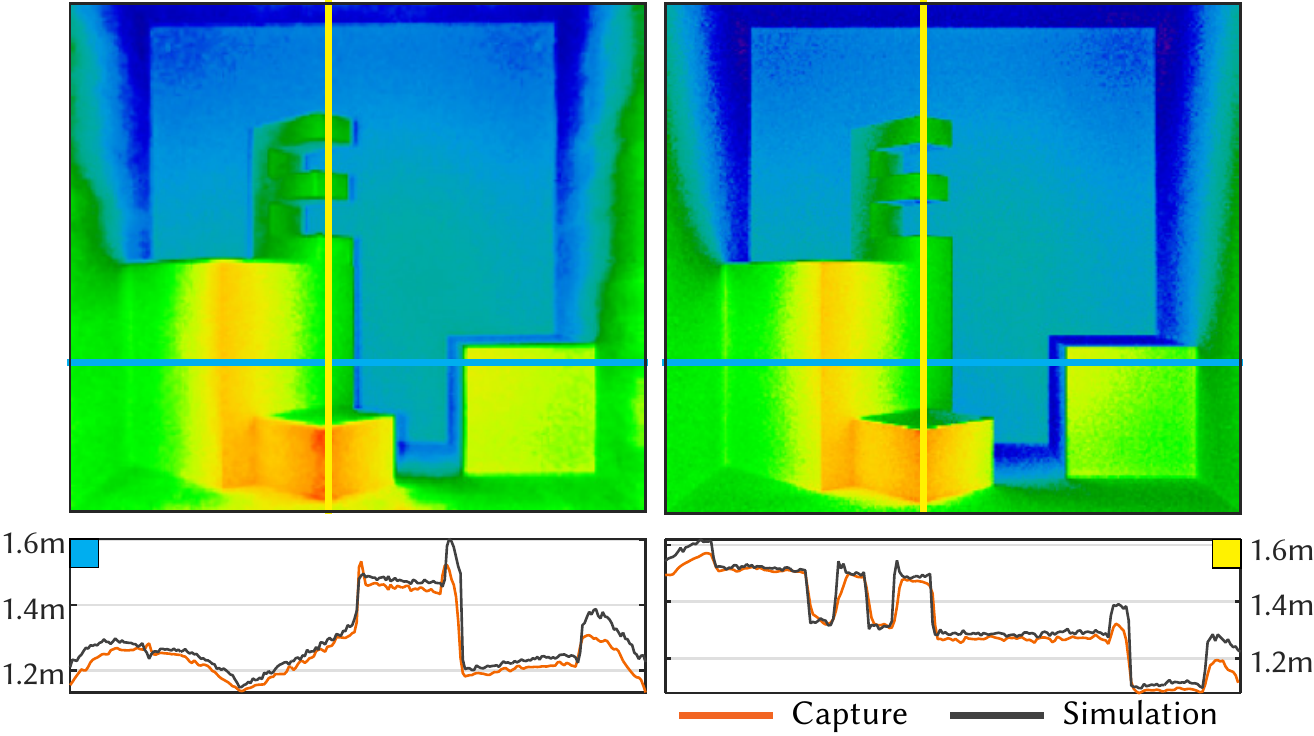}
	\caption[]{Depth map with MPI of a real scene (top left) captured by a \ToF camera, and its corresponding synthetic model (top right). The bottom row shows the depth profiles for the horizontal (left) and vertical (right) lines, showing a good agreement.}
	\label{fig:cap_vs_sim}
\end{figure}

\Fig{cap_vs_sim} compares a real scene captured with a \ToF camera (thus including MPI) with our \ToF simulation, showing a good match. Additionally, in \Apx{depth_stats} we perform a statistical analysis on both datasets, to assess their similarity.

\paragraph*{Quantitative analysis of MPI errors}
\begin{figure}[t]
	\centering
	\includegraphics[width=\columnwidth, page=2]{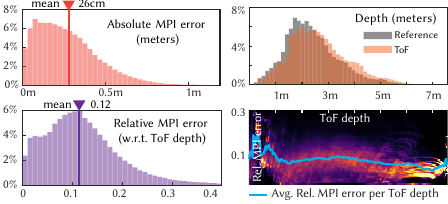}
	\caption{In reading order: Absolute MPI error; depth distribution for reference and \ToF depths, at a modulation of 20MHz (i.e. maximum unambiguous distance of 7.5m); relative error with respect to measured \ToF depth; and bivariate histogram showing relative MPI error density per measured \ToF depth. Measured \ToF depths contain an average error of 12\%. The blue line in the bivariate histogram shows that the average relative MPI error per \ToF depth remains around 10\% for most measured depths.}
	\label{fig:tof_ref_stats}
\end{figure}
\Fig{tof_ref_stats} shows the error distributions across our entire synthetic dataset. Note that while previous works have addressed local errors of just a few centimeters in small scenes, our data indicates that the global component can introduce much larger errors, with an average error of 26 cm (red line in the top-left histogram) for scenes up to 7.5 m. On average, 12\% of the measured \ToF depth corresponds to MPI (bottom-left histogram). The bivariate histogram relating relative error and observed depth (bottom-right) additionally shows that the average remains constant at around 10\% for most measured depths, being larger for smaller depths.

%% file: 50_learning_design.tex
\section{Network Architecture }
\label{sec:learning}

\begin{figure*}[t]
	\includegraphics[width=0.9\textwidth,page=1]{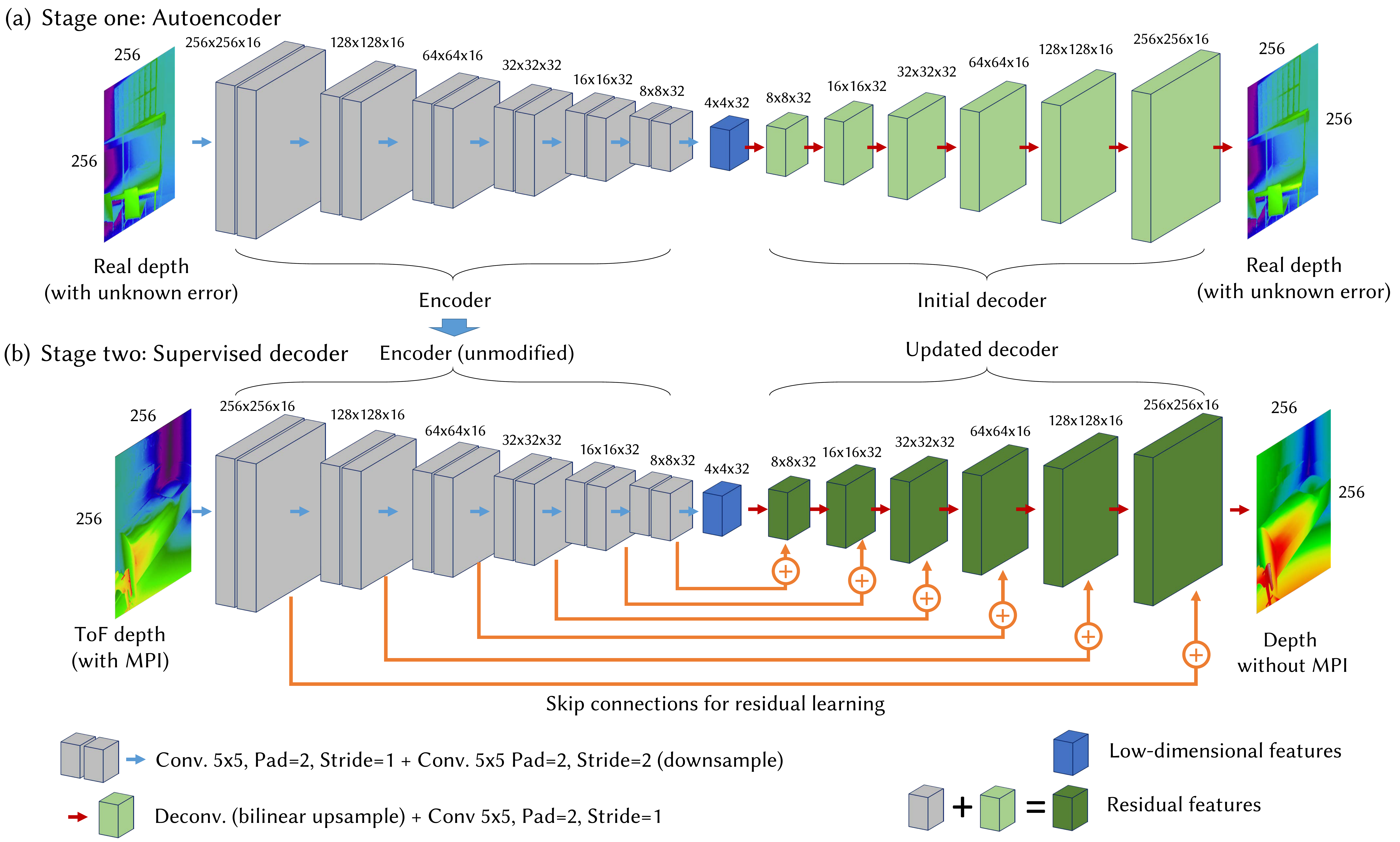}%
	\caption[]{Our two-stage training process for our regression network architecture. The first training stage learns a convolutional autocoder (an encoder-decoder pair) to learn lower-dimensional representations of depth. We use incorrect (with MPI) depth as both input and output in this unsupervised training. 
		In the second, supervised training stage we input both incorrect depth with MPI and depth without MPI, and add skip connections for residual learning while fixing the weights of the encoder. 
		This supervised training allows us to update the weights of a reconstruction decoder.
		Our final network is the pair of the original encoder and the updated decoder, working as a regression network. Please refer to the text for a more complete description.}
	\label{fig:net_arch}
\end{figure*}

We now describe how to train our network, following our two-stage scheme with the real dataset (during autoencoding) and our synthetic dataset (during supervised decoding). 

\subsection{Stage One: Autoencoder}
\label{sec:cae_learning}

%
We train our convolutional autoencoder using the real dataset, containing 48000 depth images with unknown errors.
We use this incorrect depth as input and output for this unsupervised training, and use the synthetic dataset (with MPI) as its validation set.
%
%
With this stage we pre-initialize the network so the encoder (\Fig{net_arch}, top, gray blocks) is able to generate lower-dimensional feature vectors (\Fig{net_arch}, top, blue block) for both real and synthetic depth maps. Training and validation curves of this stage are shown in \Fig{learning_curves}. 
Once we train the parameters of the encoder, we freeze its convolutional layers and update only the decoder layers in the second stage.


\paragraph{Network Parameters}
In the encoding stage, we apply sets of two 5\x5 convolutions with a two-pixel padding, and a stride of two pixels to progressively reduce the size of the convolutional inputs to each layer. 
This helps to effectively combine and find features at different scales. We perform this operation at six scales, applying pairs of convolutions over features of 256\x256 pixels (input), down to 8\x8 (innermost convolution pair, last encoding layer). In the decoding stage, we perform upsampling and 5\x5 convolutions with two-pixel padding, starting from the encoder output (see \Fig{net_arch} top, red arrows) until we reach the  output resolution 256\x256.

We have tested our network without this pre-initialization step, feeding it directly with synthetic labeled data. The results show that the network loses the ability to generalize, arbitrarily decreasing the accuracy in the validation dataset, as presented in \Sec{results}.

\subsection{Stage Two: Supervised Decoder}
\label{sec:residualnet}
\begin{figure}[t]
	\centering
	\includegraphics[width=\columnwidth, page=2]{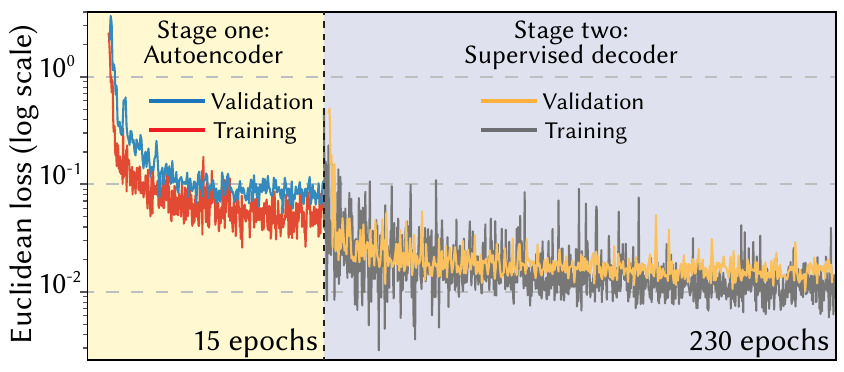}
	\caption{Learning curves for our two-stage scheme. The first stage (left) learns lower-dimensional representations of real depths using an autoencoder. The second stage (right) learns MPI corrections in the decoder by training with our labeled synthetic data. Note that the first stage converges quickly and provides a good starting point for the second stage, which uses a different training set.}
	\label{fig:learning_curves}
\end{figure}

In the second stage, we freeze the encoder layers, and train the decoder through supervised learning using our synthetic dataset. We introduce incorrect depth (with MPI) as input, and target reference depth (without MPI) as output. We use 80\% of our synthetic dataset for training, and the remaining 20\% for validation. 
%
Given that the encoder performs downsampling operations to detect features at multiple scale levels, full resolution outputs (256\x256) are significantly blurred. 
We thus add symmetric skip connections to mix detailed features of the encoding convolutions (which stay unchanged) to their symmetric outputs in the decoder (\Fig{net_arch}, bottom). Since we observed that the difference between depth with MPI (input) and reference depth (output) is on average 12\%, we treat MPI as a residue~\cite{he2016deep} by performing element-wise additions between the upsampled features and the skipped ones. Training and validation curves of this stage are shown in \Fig{learning_curves}.

In principle, concatenation of skipped features (instead of simpler element-wise additions) could create more complex combinations with upsampled features using additional learned filters. However, in our results we observed that our residual approach performs equally well (even slightly better, see \Sec{alt_nets}) while yielding a 30\% smaller network model, reducing also execution time. 

\subsection{Implementation Details}
\review{
We have implemented our network in Caffe, and trained it on an NVIDIA GTX 1080. Our network takes the input depth without applying any normalization. Following previous works using CNNs, all convolutional layers are followed by a batch normalization layer, a scale and bias layers, and a ReLU activation layer, in that order. For training, we use the Adam solver~\cite{kingma2014adam} for gradient propagation. The learning rate was set to $1\cdot10^{-4}$, and adjusted in a stepped fashion in steps between $1\cdot10^{-3}$ and $1\cdot10^{-5}$, to avoid getting stuck in a plateau, while our batch size is set to 16 to maximize memory usage. 
Our resulting network performs MPI corrections for a single frame in 10 milliseconds.
Additional details on the definition of both the network and training, including the input sources, can be found in the project webpage. }

%% file: 60_results.tex

\section{Results and Validation}
\label{sec:results}

In this section, we first analyze other alternative, simpler networks, showing how they yield inferior results. We then compare our results against existing methods using off-the-shelf cameras, and thoroughly validate the performance of our approach in both synthetic and real scenes, including video in real time.  
Real scenes were captured with a PMD CamCube 3.0, which provides depth images at 200\x200 resolution, and operates at 20MHz with 7.5 meters of maximum unambiguous depth. 
%

\subsection{Alternative Networks}
\label{sec:alt_nets}

\begin{figure}[t]
	\centering
	\includegraphics[width=\columnwidth, page=2]{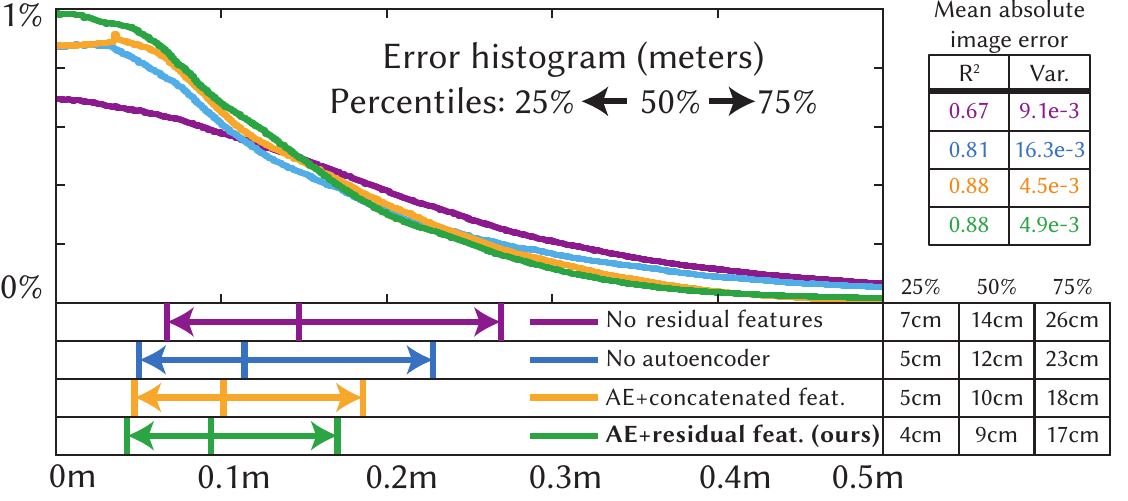}
	\caption{Error distributions for different network alternatives: without residual connections (purple),  without the autoencoding pre-initialization (blue), with concatenated skip connections (yellow), and our residual approach (green). Average R$^2$ across all predicted depth images shows 
		that our residual learning with autoencoding pre-initialization reaches the best error distributions in the results, in terms of accuracy and low variance. The percentiles show that our approach presents also the best error distribution.
	}
	\label{fig:net_alternatives_hist}
\end{figure}

We test three alternatives to our CNN: (1) suppressing the pre-initialization autoencoder stage by directly training an encoder-decoder with synthetic labeled data; (2) removing the residual skip connections; (3) substituting residual connections by concatenated connections. \Fig{net_alternatives_hist} shows how our autoencoder with the residual learning approach yields better results with respect to these other alternatives, with better generalization and smaller network size.
By computing $\textrm{R}^2$ scores between the depth predicted by each network and the target depths across the whole set of images, 
we observe that, without the autoencoding stage, the images present a lower average score than our results (see \Fig{net_alternatives_hist}, top-right table). Also, the variance of the per-image mean absolute error without pre-initialization triplicates the variance of our residual network errors, leading to more unstable accuracy.
Concatenating skip connections worsens results slightly, while additionally  making the network about 30\% larger, due to the need to learn more parameters to combine the additional features. 
Last, removing residual skip connections avoids enriching upsampled features with high resolution features from the encoder layers, producing blurry outputs and therefore a much higher error. 

\input{63_comparison_previous_work}

\subsection{Synthetic Scenes}
\begin{figure}[t]
	\centering
	\includegraphics[width=\columnwidth, page=1]{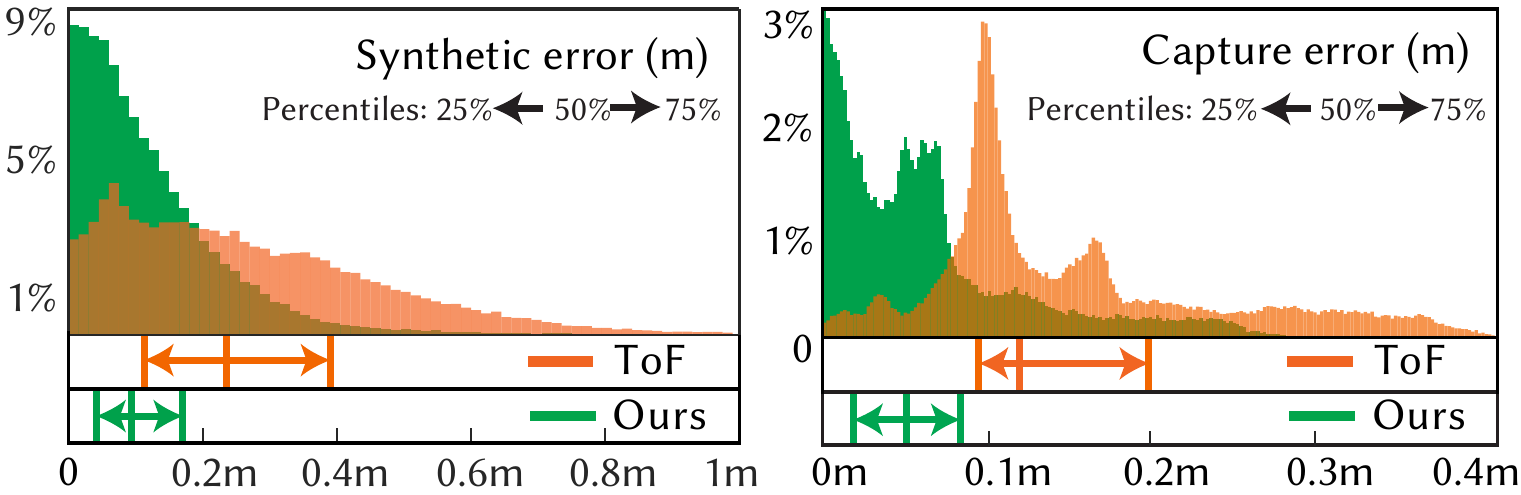}
	\caption{Per-pixel distributions of absolute error for the synthetic validation dataset (left), and real dataset with measured ground truth (right). For each distribution, three percentiles ($25\%$, $50\%$ and $75\%$) are marked below. Our results clearly present a better error distribution. 
	}
	\label{fig:results_synth_validation_1_16_hist_curve}
\end{figure}
\begin{figure*}[t]
	\centering
	\includegraphics[width=\textwidth, page=1]{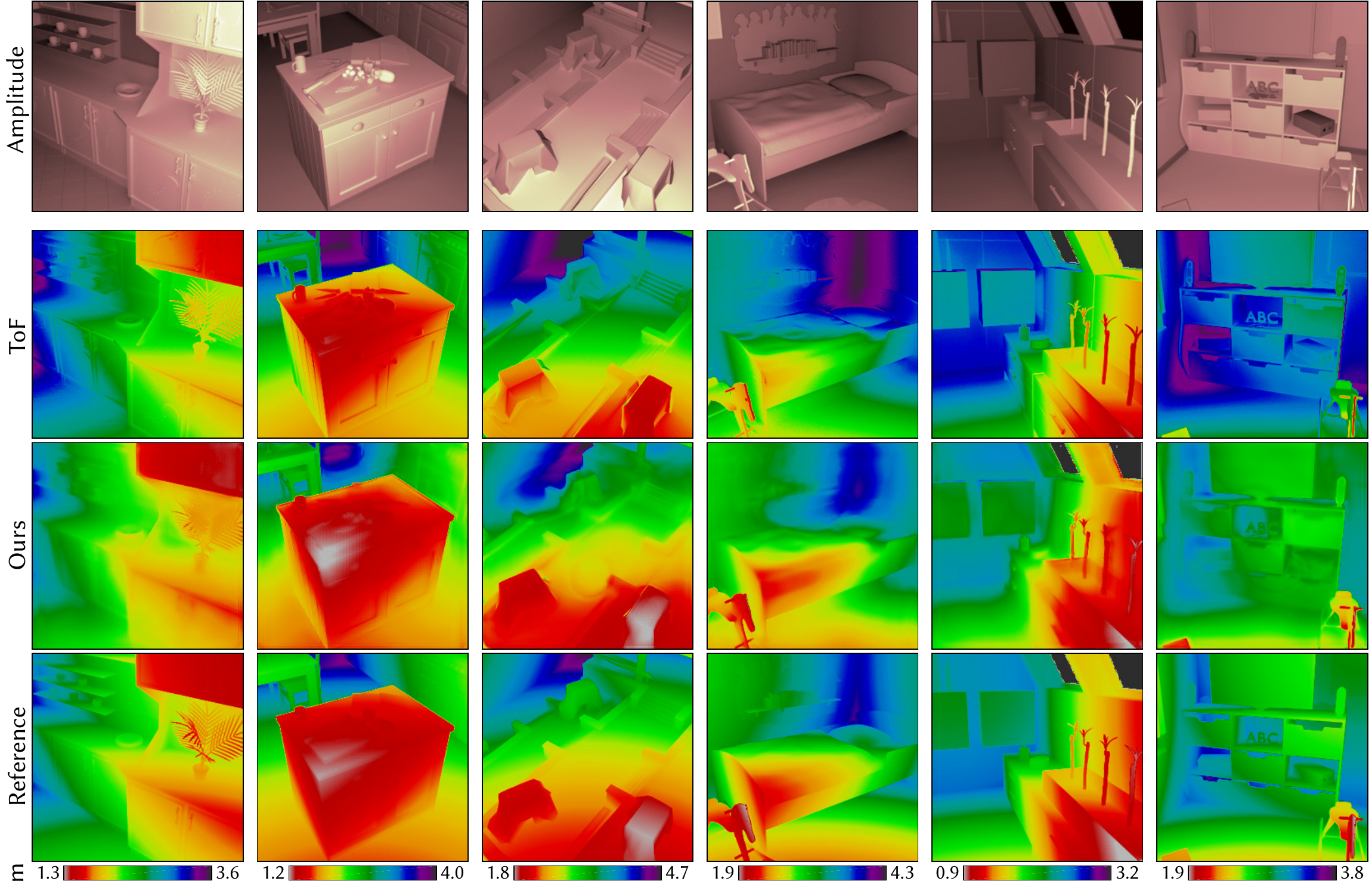}
	\caption{Validation results for synthetic scenes (validation set) at varying distances between 0.5 and 7 meters. Top row shows \ToF amplitude. Second, third and fourth rows show \ToF depth with MPI, our estimated depth, and reference depth (without MPI). Our solution manages to correct MPI errors in a wide range of scenes while preserving details.}
	\label{fig:results_synth_validation_1_16}
\end{figure*}
\begin{figure*}[t]
	\centering
	\includegraphics[width=\textwidth, page=1]{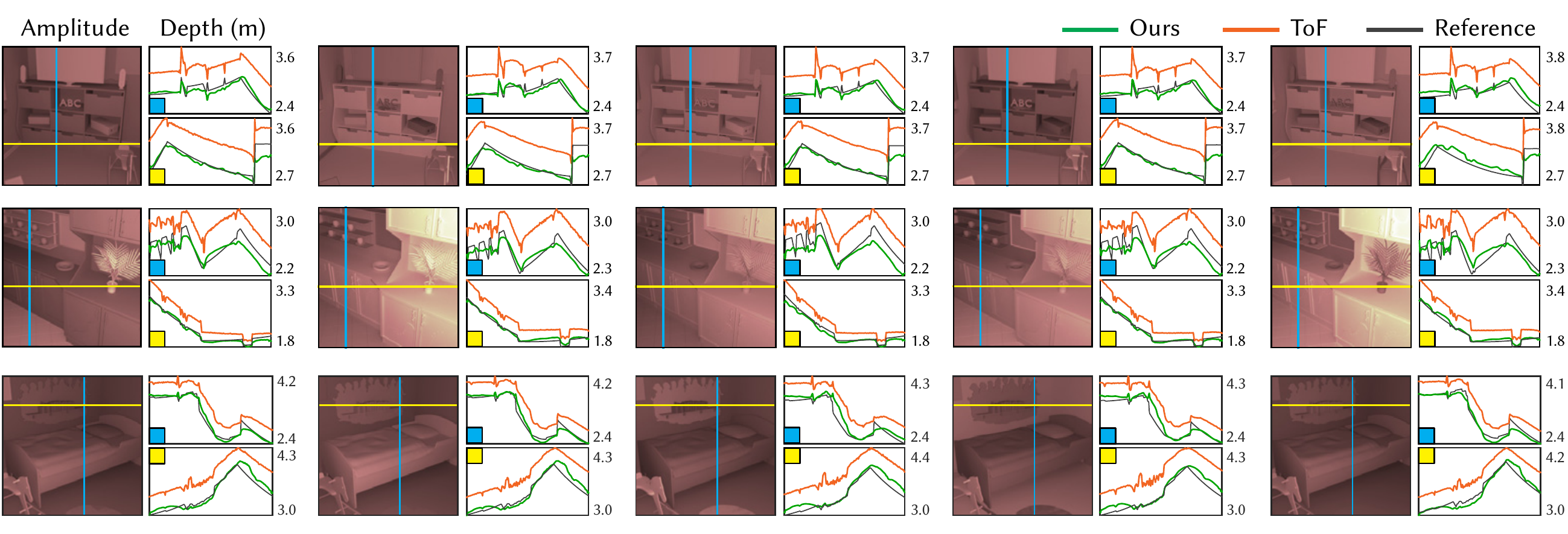}
	\caption{Validation results for three  scenes (one per row) with randomly varying albedos (one albedo set per column). Images show amplitude, vertical and horizontal plots show depth. Our approach is robust under arbitrary albedo variations.}
	\label{fig:results_synth_valbedo}
\end{figure*}

From the synthetic dataset, a total of 213 scenes were used for the validation set (augmented to 1704 with flips and rotations). 
As \Fig{results_synth_validation_1_16_hist_curve} (left) shows, our method yields a much better error distribution. 
\Fig{results_synth_validation_1_16} shows a comparison of simulated \ToF depth, our MPI-corrected depth, and the reference depth images. Our CNN preserves details while significantly mitigating MPI errors. 

Additionally, in \Fig{results_synth_valbedo} we compare the errors for five albedo combinations of three different scenes by randomly varying each object's reflectance between 0.3 and 0.8. It can be observed how our network is robust to these variations, consistently correcting depth errors due to MPI. Note how even under strong albedo changes on large flat objects (e.g., the cabinet in the first row, or the tabletop in the second) our network successfully recovers the correct depth. 

\subsection{Real Scenes}
\begin{figure}[h]
	\centering
	\includegraphics[width=\columnwidth, page=2]{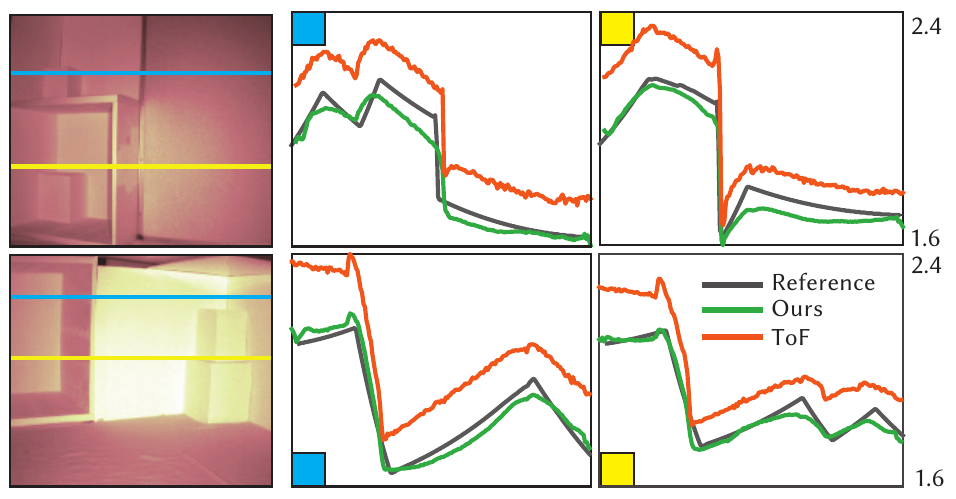}
	\caption{Error comparison for different captured combinations of the Cornell box, panels and prisms. Reference depth solutions were obtained replicating the scenes in simulation. We significantly decrease MPI errors in all the captured scenes, yielding errors under 5 cm for the 50\% of the pixels, as we demonstrate in the error histograms (\Fig{results_synth_validation_1_16_hist_curve}).}
	\label{fig:results_cornell}
\end{figure}
We now analyze the performance of our method in real scenes, captured with a PMD camera. We first show results on controlled scenarios with combinations of V-shapes, panels and a Cornell box, and then more challenging captures in the wild. The lens distortion of all the captures was corrected using a standard calibration of the intrinsic camera parameters, using a checkerboard pattern and captures at different distances \cite{lindner2010time,bouguet2004camera}.

\paragraph*{Cornell box and V-shapes}
We created different setups combining a Cornell box structure and V-shapes with flat panels (see \Fig{results_cornell}, left column). We accurately measured the geometry of these scenes, to create corresponding synthetic reference images for a quantitative analysis.
The Cornell box dimensions were 600\x500\x640 mm, with additional panels from 400 mm to 1200 mm.
%
The PMD camera was placed at multiple distances from 0.5 to 2.4 meters. 
We added several geometric elements to the scenes: three prisms with different dimensions, and a cardboard letter E, in order to add extra sources of MPI error. The surfaces of the Cornell box, two panels, and the smaller shapes were painted twice with a 50-50 mixture of barium sulfate and white matte paint, providing a good trade-off between durability and high-reflectance diffuse surfaces \cite{knighton2005mixture,patterson1977kubelka}. Note that this mixture has an albedo of approximately 0.85, leading to large MPI and thus ensuring very challenging scenarios.
\Fig{results_cornell} (middle and right) shows the results of the captured depth and our corrected result. We compensate most of the MPI errors in both scenes, approximating depth much closer to the reference than the \ToF camera. 
We can observe on the error distributions (right histogram in \Fig{results_synth_validation_1_16_hist_curve}) how our CNN manages to keep 50\% of the per-pixel errors under 5 cm, while 75\% of the errors in the PMD captures are \emph{over} 9 cm.

\paragraph*{Scenes in the wild}
\begin{figure}[t]
	\centering
	\includegraphics[width=\columnwidth]{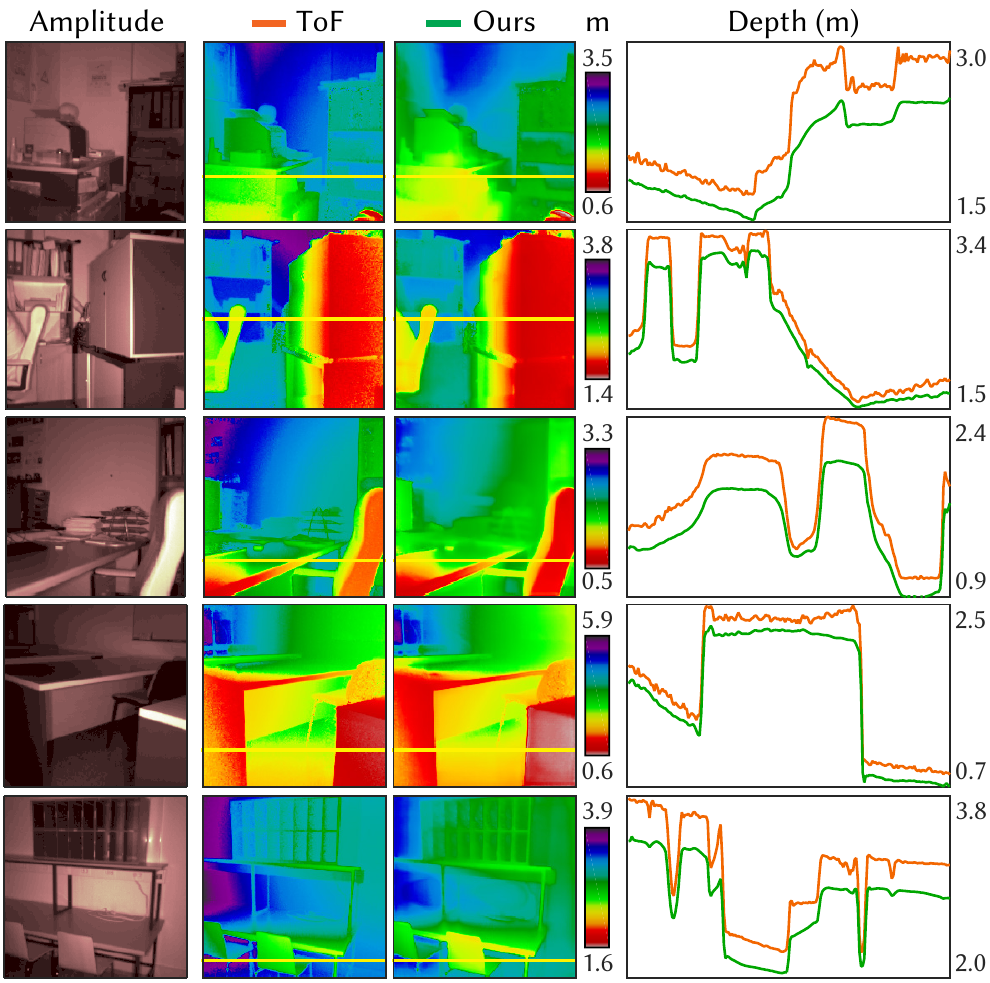}
	\caption{Comparison of conventional \ToF and our corrected depth, in real world scenes. In accordance with our analysis of MPI errors from our synthetic dataset, longer depths yield higher errors (about 10\% of the measured distance).}
	\label{fig:results_real_scenes}
\end{figure}
We now analyze several in-the-wild scenes, to illustrate the benefits of our approach in non-controlled conditions. 
The results are shown in \Fig{results_real_scenes}. We can see how our network successfully suppresses MPI in all cases, while still preserving details thanks to our residual learning approach. The magnitude of our MPI corrections is proportional to the measured \ToF distance. This follows our observations in the error analysis (see \Fig{tof_ref_stats}), where larger distances tend to yield larger errors since the relative error oscillates at around 10\%.

\subsection{Video in Real Time}
Given the speed of our approach we are also able to process depth videos in real time. Regarding temporal coherence, we leverage the fact that our input (incorrect depth) is quite stable between frames, so our network produces temporally coherent results without explicitly enforcing it. 
Other inputs such as amplitude and/or phase could show less stability, compromising this temporal coherence. We show some frames in \Figs{teaser}{results_videos}. Full sequences can be found in the supplemental video.

\begin{figure}[t]
	\centering
	\includegraphics[width=\columnwidth, page=1]{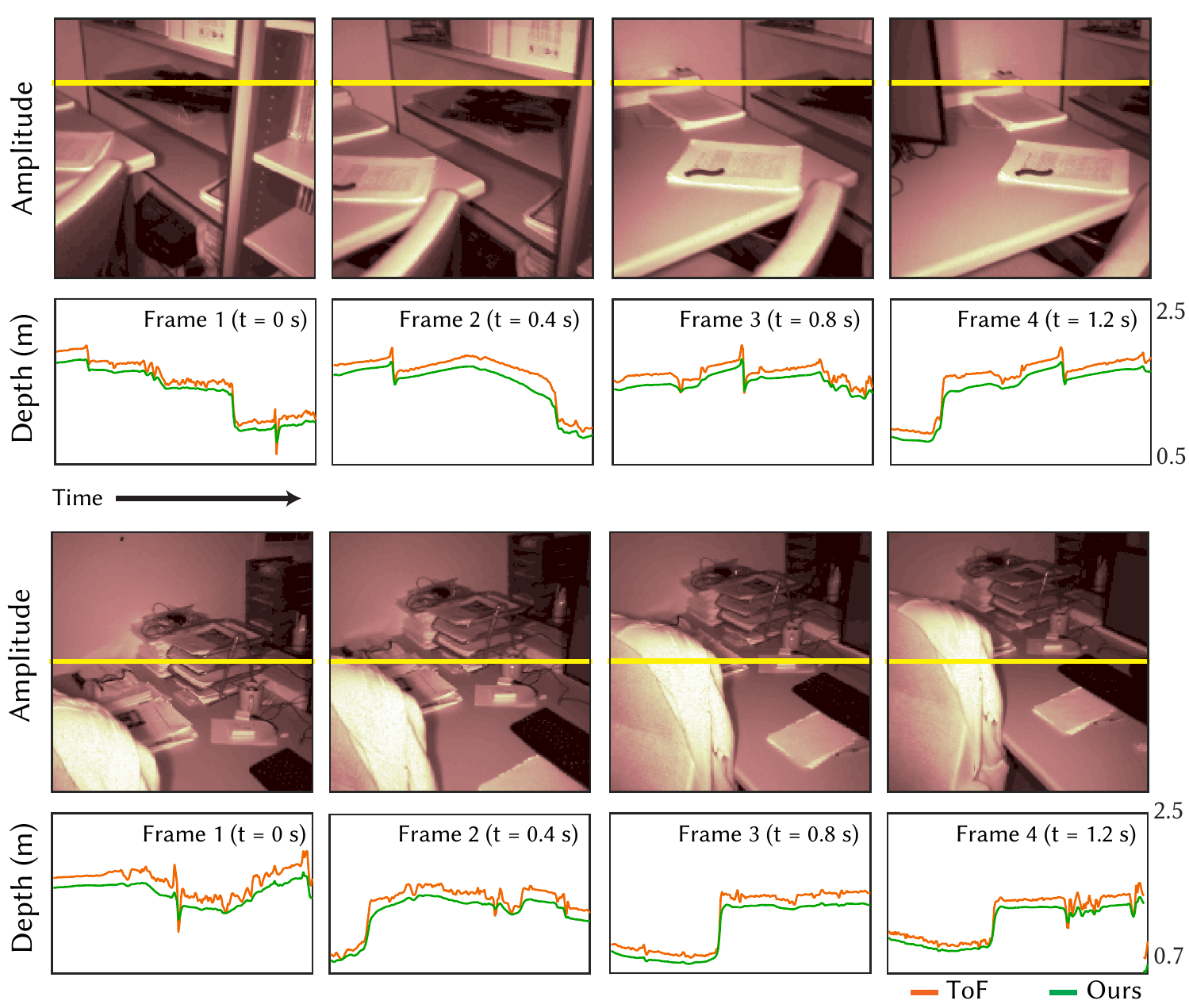}
	\caption{Our approach can also be applied to correct MPI errors of in-the-wild videos of real scenes, in real time and keeping temporal consistency. Here we show the depth profiles of a few frames of two of our videos. The complete sequences can be found in the supplemental video.}
	\label{fig:results_videos}
\end{figure}

%% file: 63_comparison_previous_work.tex
\subsection{Comparison with Previous Work}
\begin{figure*}[t]
	\centering
	\includegraphics[width=\textwidth, page=1]{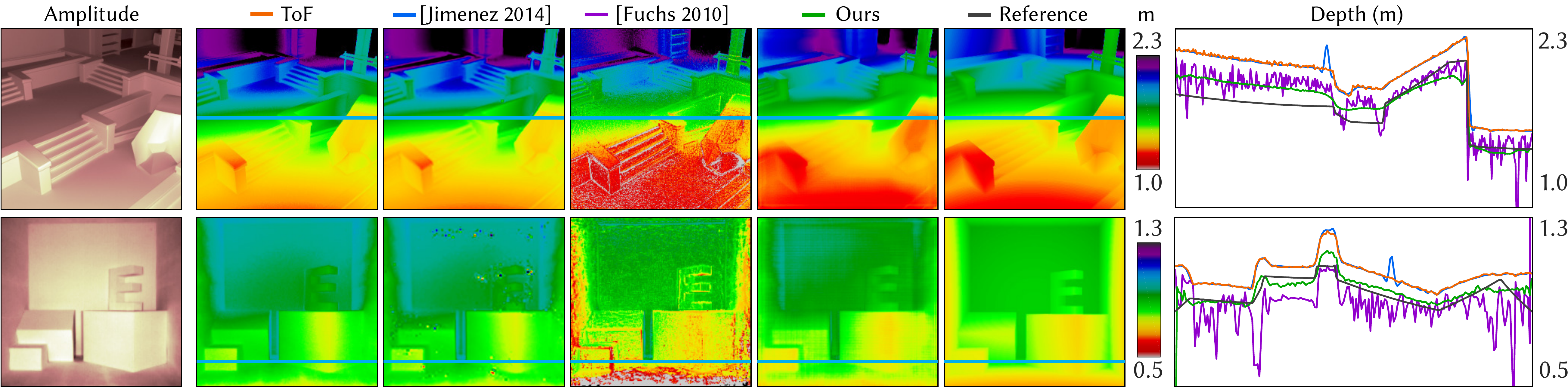}
	\caption{Comparison with previous works. Top row: synthetic scene. Bottom row: real  scene. Jimenez et. al' approach~\shortcite{jimenez2014modeling} is hindered by the geometrical complexity of the scenes, and fails to correct MPI since the optimization converges to a local minimum. Fuchs' technique~\shortcite{fuchs2010multipath} is closer to the reference, but is very noisy and greatly diverges in some regions. Our approach is the closest to the reference, as shown both in the depth maps and the graphs plotting the blue scanlines.}
	\label{fig:comparison_previous_work}
\end{figure*}
In \Fig{comparison_previous_work} we compare our solution to previous works requiring no hardware modifications, and using a single frequency \cite{fuchs2010multipath,jimenez2014modeling}. We use both a synthetic and a real scene. 
%
%
Fuchs' approach~\shortcite{fuchs2010multipath} results in a noisy estimation due to the discretization of light transport, taking around $10$ minutes to compute. Jimenez et. al's technique~\shortcite{jimenez2014modeling} is hindered by the geometrical complexity of the scenes, taking around one hour for a lower resolution image of $100\times100$; we were unable to compute larger images due to high memory consumption (about 60GB). Moreover, the results are very close in general to the input \ToF captures, including MPI errors and several outlier pixels.
Our results are significantly closer to the reference, eliminating MPI errors, while being orders of magnitude faster. 



%% file: 70_conclusion.tex
\section{Discussion and future work}
\begin{figure}[t]
	\centering
	\includegraphics[width=\columnwidth, page=3]{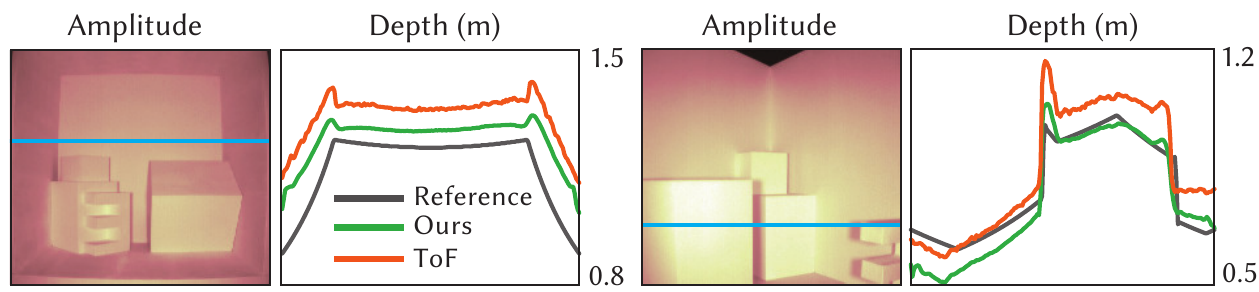}
	\caption{Due to the high albedo of our barium-mixed diffuse paint  ($\approx 0.85$), some specific camera-light configurations may yield large MPI errors (left). Objects very close to the camera (right, bottom-left box) yield a higher error since most of our training dataset has depths from 1.5m to 4m (see \Fig{tof_ref_stats}). Still, our approach manages to improve MPI errors in both cases, providing results significantly closer to the reference depth.}
	\label{fig:failure_cases}
\end{figure}
We have presented a new approach for  \ToF imaging, to compensate the effect of multipath interference in real time, using an unmodified, off-the-shelf camera with a single frequency, and just the incorrect depth map as input. This is possible due to our carefully designed encoder-decoder (convolutional-deconvolutional) neural network, with a two-stage training process both from captured and synthetic data. With this paper, we provide our synthetic time-of-flight dataset that includes pairs of incorrect depth (affected by MPI) and its corresponding correct depth maps, as well as the trained network, for public use. 

Several avenues of future work exist. 
\review{First, as discussed in \Sec{datageneration}, we do not consider the wiggling error due to non-perfectly sinusoidal waves in our training dataset, since it is partially compensated by ToF cameras, and manufacturers do not provide information on this. If this information were available, we could incorporate the full camera pipeline (including non-sinusoidal waves and wiggling correction) into our training dataset, and re-train our CNN accounting for these residual errors. } 
In addition, there are still challenging scenarios where results could be improved, as shown in \Fig{failure_cases}. The very high albedo of barium-painted surfaces creates large MPI errors, specially under specific camera-light configurations (left). Although our MPI correction provides better results than the captured \ToF depth, there is still some residual error of about 10 cm in average. Also, our network fails to correct MPI in the presence of objects which are very close to the camera, such as the bottom-left box in the second example (\Fig{failure_cases}, right). This is most likely because most of our synthetic dataset contains depths between 1.5m and 4m. Despite this, as we showed in \Fig{results_synth_validation_1_16_hist_curve}, per-pixel error distributions are significantly better than captured \ToF depth. 

\review{Our work assumes diffuse (or nearly diffuse) reflectance. Although we have shown that it works well in several real-world scenarios with more general reflectances, it presents some problems in the presence of highly glossy materials. While incorporating such reflectances into our training dataset would help, our approach is likely to fail for extremely glossy or transparent surfaces; in such scenarios, other multi-frequency approaches~\cite{kadambi2013coded,Qiao2015} could be better suited.  }
%

%% file: 99_acks.tex

\section*{Acknowledgements}
We want to thank the anonymous reviewers for their insightful comments, Belen Masia for proofreading the manuscript, and the members of the Graphics \& Imaging Lab for helpful discussions. 
This project has received funding from the European Research Council (ERC) under the European Union's Horizon 2020 research and innovation programme (CHAMELEON project, grant agreement No 682080), DARPA (project REVEAL), and the Spanish Ministerio de Econom\'{i}a y Competitividad (projects TIN2016-78753-P and TIN2014-61696-EXP). Min H. Kim acknowledges Korea NRF grants (2016R1A2B2013031, 2013M3A6A6073718), Giga KOREA Project (GK17P0200) and KOCCA in MCST of Korea. Julio Marco was additionally funded by a grant from the Gobierno de Arag\'{o}n.

%% file: 80_light_transport.tex
\section*{Appendices}
\section{Light transport in image space}
\label{ap:light_transport}

\Sec{problem} shows how most of the information on multipath interference from a scene is available in image space. This allows us to approximate \Eq{pixelmultiple} by limiting the integration domain $\pixelpathrange$ to the differential paths $\pixelpath$ that reach pixel $\pixel$ from visible geometry. 
%
%
Moreover, given the discretized domain of an image, we can model \Eq{pixelmultiple} using the transport matrix $\sTransportMatrix_i$~\cite{Ng2003,otoole2014probing}, which relates the ideal response in pixel $\pixel_v$ with the outgoing response at pixel $\pixel_u$, for a measurement phase shift $i$. Thus 
\begin{align}
\sCorrApp_{i}(\pixel_u) & = \sCorr_{i}(\pixel_u) + \sum_v 
\sTransportMatrix_i(\pixel_u,\pixel_v)\,\sCorr_{i}(\pixel_v)
\nonumber \\
&= \sCorr_{i}(\pixel_u) + \sTransportMatrix_i \ast \sCorr_i,
\label{eq:pixelmultiple_transport}
\end{align}
where $\ast$ is the convolution in $\pixel_v$, and $\sCorr_i$ is the full phase-shifted image. 
While ideally this means that we can compute the correct phase-shifted image $\sCorr_i$ by applying a \textit{\textbf}{deconvolution} on the captured $\sCorrApp_{i}$, as $\sCorr_i = \sCorrApp_i \ast_v (I+\sTransportMatrix_i)^{-1}$ with $I$ the identity matrix, in practice this is not possible since the transport matrix $\sTransportMatrix_i$ is unknown. Capturing it is an expensive process, and we cannot make strong simplifying assumptions on the locality of light transport (i.e. sparsity in $\sTransportMatrix_i$), since light reflected from far away pixels might have an important contribution on pixel $\pixel$. \review{However, as we show in \Sec{approach}, we can learn the resulting deconvolution operator by means of a convolutional neural network.}

%% file: 90_depth_stats.tex

\section{Depth statistics}
\label{ap:depth_stats}
To validate our synthetic dataset (\Sec{datageneration}),
we follow previous works on depth image statistics~\cite{huang1999statistics,huang2000statistics} on both real and synthetic datasets. In particular, we analyze single-pixel, derivative and bivariate statistics, as well as joint statistics of Haar wavelet coefficients. To compare the results we use three different metrics: chi-squared error~\cite{pele2010quadratic}, the Jensen-Shannon distance~\cite{lin1991divergence,endres2003new}, and Pearson's correlation coefficient. 
The first is a weighted Euclidean error ranging from 0 to $\infty$ (less is better); the second one measures the similarity between two distributions, ranging from 0 to $\sqrt{ln(2)}=0.833$ (less is better); the last measures correlation, where a value of 0 indicates two independent variables, and $\pm 1$ indicates a perfect linear direct or inverse relationship. 

\begin{figure}[t]
	\centering
	\includegraphics[width=\columnwidth]{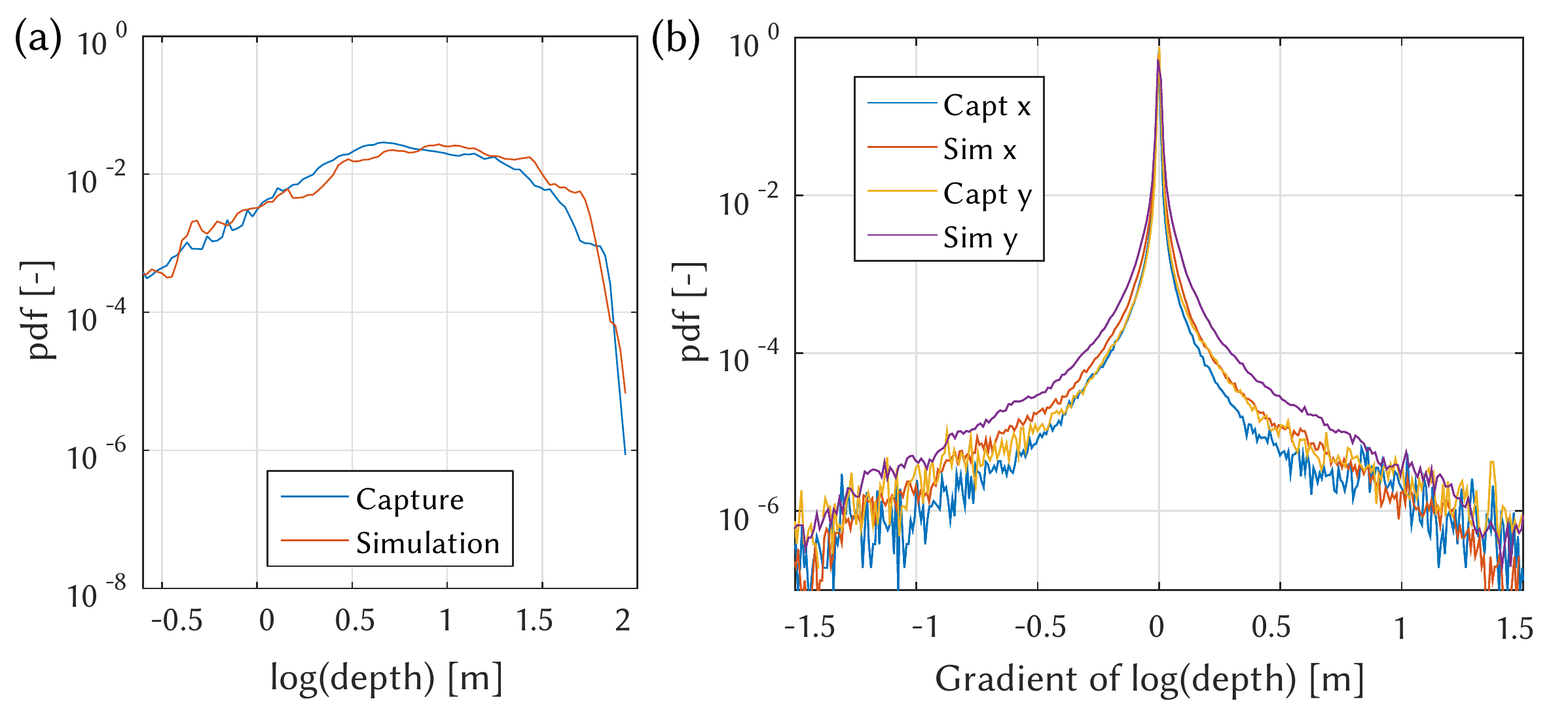}%
	\vspace{-2mm}%
	\caption[]{%
	(a) Logarithmic depth histogram of real and simulated depths, showing a good match between both sets. (b) Derivative in x and y directions of real and simulated depths, also showing a similar trend. Please refer to the text for quantitative data and other statistical analyses. 
	}
	\label{fig:combined}
	\vspace{-4mm}
\end{figure}

\Fig{combined}a shows the depth histograms of both sets of images. They are very similar, with a chi-squared error of 0.032, Jensen-Shannon distance of 0.129, and a correlation of 0.90. 
%
\Fig{combined}b  shows how the vertical and horizontal gradients of both datasets also follow a similar trend, with a chi-squared error of 0.051 and 0.028, Jensen-Shannon distance of 0.162 and 0.118, and correlation coefficients of 0.948 and 0.969 for the vertical and horizontal gradients respectively. Both sets have high kurtosis values, as reported by Huang \shortcite{huang2000statistics}.

Next, we carried out a bivariate statistical analysis to pairs of pixels at a fixed separation distance, following the co-ocurrence equation~\cite{lee2000random,huang2000statistics}. 
 Capture and simulation match with a chi-squared error of 0.094, Jensen-Shannon distance of 0.245, and correlation of 0.915.
Last, we selected three Haar filters (horizontal, vertical, and diagonal) in order to analyze joint statistics in the wavelet domain. 
Capture and simulation match with a chi-squared error of 0.152 and 0.172,  Jensen-Shannon distance of 0.314 and 0.336, and correlation coefficients of 0.892 and 0.896, for the horizontal-vertical and horizontal-diagonal pairs respectively. 

These values indicate that our simulated images share very similar depth statistics with existing real-world depth images, so they can be used  as reliable input for our network. 